\documentstyle[twocolumn,aps,floats,amssymb,epsfig]{revtex}

\begin{document}

\def\d{{\rm d}}
\def\e{{\rm e}}
\def\i{{\rm i}}
\def\O{{\rm O}}
\def\half{\mbox{$\frac12$}}
\def\eref#1{(\protect\ref{#1})}
\def\etal{{\it{}et~al.}}
\def\Li{\mathop{\rm Li}}
\def\av#1{\left\langle#1\right\rangle}
\def\set#1{\left\lbrace#1\right\rbrace}
\def\stirling#1#2{\Bigl\lbrace{#1\atop#2}\Bigr\rbrace}
\def\cG{{\mathcal{G}}}

\draft
\tolerance = 10000

\renewcommand{\topfraction}{0.9}
\renewcommand{\textfraction}{0.1}
\renewcommand{\floatpagefraction}{0.9}
\setlength{\tabcolsep}{4pt}

\twocolumn[\hsize\textwidth\columnwidth\hsize\csname @twocolumnfalse\endcsname

\title{Random graphs with arbitrary degree distributions and their
applications}
\author{M. E. J. Newman$^{1,2}$, S. H. Strogatz$^{2,3}$,
and D. J. Watts$^{1,4}$}
\address{$^1$Santa Fe Institute, 1399 Hyde Park Road, Santa Fe, NM 87501}
\address{$^2$Center for Applied Mathematics, Cornell University, Ithaca NY
14853--3401}
\address{$^3$Department of Theoretical and Applied Mechanics,
Cornell University, Ithaca NY 14853--1503}
\address{$^4$Department of Sociology, Columbia University,
1180 Amsterdam Avenue, New York, NY 10027}
\maketitle

\begin{abstract}
  Recent work on the structure of social networks and the internet has
  focussed attention on graphs with distributions of vertex degree that are
  significantly different from the Poisson degree distributions that have
  been widely studied in the past.  In this paper we develop in detail the
  theory of random graphs with arbitrary degree distributions.  In addition
  to simple undirected, unipartite graphs, we examine the properties of
  directed and bipartite graphs.  Among other results, we derive exact
  expressions for the position of the phase transition at which a giant
  component first forms, the mean component size, the size of the giant
  component if there is one, the mean number of vertices a certain distance
  away from a randomly chosen vertex, and the average vertex--vertex
  distance within a graph.  We apply our theory to some real-world graphs,
  including the world-wide web and collaboration graphs of scientists and
  Fortune 1000 company directors.  We demonstrate that in some cases random
  graphs with appropriate distributions of vertex degree predict with
  surprising accuracy the behavior of the real world, while in others there
  is a measurable discrepancy between theory and reality, perhaps
  indicating the presence of additional social structure in the network
  that is not captured by the random graph.
\end{abstract}


]

\vspace{1cm}

\section{Introduction}
\label{intro}
\begin{figure}
\begin{center}
\psfig{figure=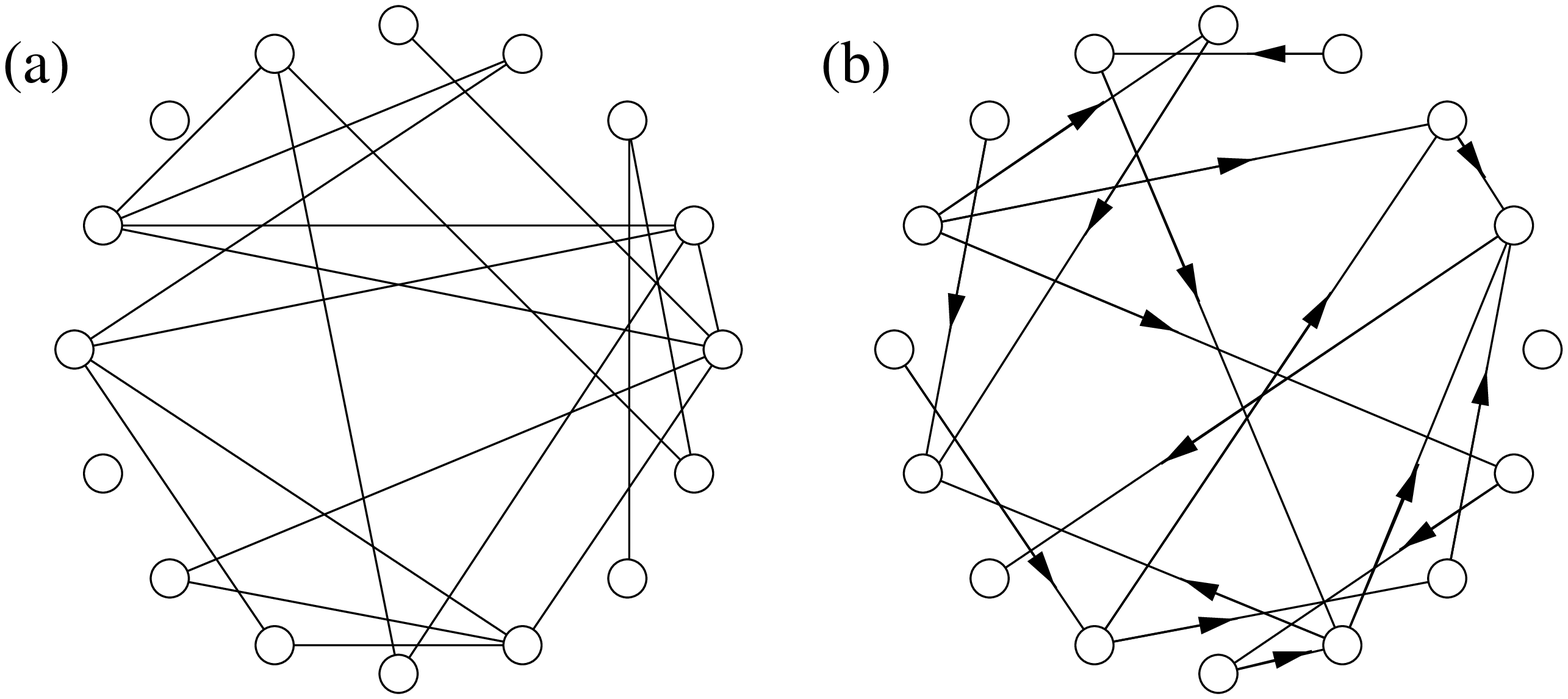,width=\columnwidth}
\end{center}
\caption{(a)~A schematic representation of a random graph, the circles
  representing vertices and the lines edges.  (b)~A directed random graph,
  i.e.,~one in which each edge runs in only one direction.}
\label{unifig}
\end{figure}

A random graph~\cite{Bollobas85} is a collection of points, or vertices,
with lines, or edges, connecting pairs of them at random
(Fig.~\ref{unifig}a).  The study of random graphs has a long history.
Starting with the influential work of Paul Erd\H{o}s and Alfr\'ed R\'enyi
in the 1950s and 1960s~\cite{ER59,ER60,ER61}, random graph theory has
developed into one of the mainstays of modern discrete mathematics, and has
produced a prodigious number of results, many of them highly ingenious,
describing statistical properties of graphs, such as distributions of
component sizes, existence and size of a giant component, and typical
vertex--vertex distances.

In almost all of these studies the assumption has been made that the
presence or absence of an edge between two vertices is independent of the
presence or absence of any other edge, so that each edge may be considered
to be present with independent probability~$p$.  If there are $N$ vertices
in a graph, and each is connected to an average of $z$ edges, then it is
trivial to show that $p=z/(N-1)$, which for large $N$ is usually
approximated by $z/N$.  The number of edges connected to any particular
vertex is called the degree $k$ of that vertex, and has a probability
distribution $p_k$ given by
\begin{equation}
p_k = \biggl( {N\atop k} \biggr) p^k (1-p)^{N-k}
    \simeq {z^k \e^{-z}\over k!},
\end{equation}
where the second equality becomes exact in the limit of large~$N$.  This
distribution we recognize as the Poisson distribution: the ordinary random
graph has a Poisson distribution of vertex degrees, a point which turns out
to be crucial, as we now explain.

Random graphs are not merely a mathematical toy; they have been employed
extensively as models of real-world networks of various types, particularly
in epidemiology.  The passage of a disease through a community depends
strongly on the pattern of contacts between those infected with the disease
and those susceptible to it.  This pattern can be depicted as a network,
with individuals represented by vertices and contacts capable of
transmitting the disease by edges.  The large class of epidemiological
models known as susceptible/infectious/recovered (or SIR)
models~\cite{SS88,AM95,KM96} makes frequent use of the so-called fully
mixed approximation, which is the assumption that contacts are random and
uncorrelated, i.e.,~that they form a random graph.

Random graphs however turn out to have severe shortcomings as models of
such real-world phenomena.  Although it is difficult to determine
experimentally the structure of the network of contacts by which a disease
is spread~\cite{Heckathorn97}, studies have been performed of other social
networks such as networks of friendships within a variety of
communities~\cite{FRO63,FS64,BKEMS98}, networks of telephone
calls~\cite{ABW98,ACL00}, airline timetables~\cite{ASBS00}, and the power
grid~\cite{WS98}, as well as networks in physical or biological systems,
including neural networks~\cite{WS98}, the structure and conformation space
of polymers~\cite{JSB00,SAB00}, metabolic pathways~\cite{FW00,JTAOB00}, and
food webs~\cite{WM00,MS00}.  It is found~\cite{ACL00,ASBS00} that the
distribution of vertex degrees in many of these networks is measurably
different from a Poisson distribution---often wildly different---and this
strongly suggests, as has been emphasized elsewhere~\cite{BA99}, that there
are features of such networks which we would miss if we were to approximate
them by an ordinary (Poisson) random graph.

Another very widely studied network is the internet, whose structure has
attracted an exceptional amount of scrutiny, academic and otherwise,
following its meteoric rise to public visibility starting in 1993.  Pages
on the world-wide web may be thought of as the vertices of a graph and the
hyperlinks between them as edges.  Empirical
studies~\cite{AJB99,HA99,KKRRT99,BKMRRSTW00} have shown that this graph has
a distribution of vertex degree which is heavily right-skewed and possesses
a fat (power-law) tail with an exponent between $-2$ and~$-3$.  (The
underlying physical structure of the internet also has a degree
distribution of this type~\cite{FFF99}.)  This distribution is very far
from Poisson, and therefore we would expect that a simple random graph
would give a very poor approximation of the structural properties of the
web.  However, the web differs from a random graph in another way also: it
is directed.  Links on the web lead from one page to another in only one
direction (see Fig.~\ref{unifig}b).  As discussed by
Broder~\etal~\cite{BKMRRSTW00} this has a significant practical effect on
the typical accessibility of one page from another, and this effect also
will not be captured by a simple (undirected) random graph model.

\begin{figure}
\begin{center}
\psfig{figure=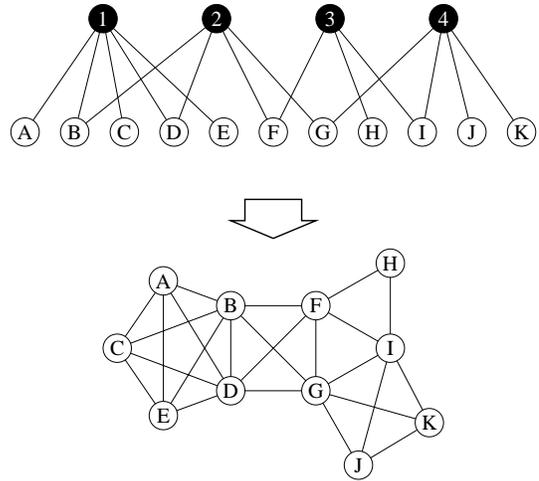,width=7cm}
\end{center}
\caption{A schematic representation (top) of a bipartite graph, such as the
  graph of movies and the actors who have appeared in them.  In this small
  graph we have four movies, labeled 1 to 4, and eleven actors, labeled A
  to K, with edges joining each movie to the actors in its cast.  In the
  lower part of the picture we show the one-mode projection of the graph
  for the eleven actors.}
\label{bifig}
\end{figure}

A further class of networks that has attracted scrutiny is the class of
collaboration networks.  Examples of such networks include the boards of
directors of companies~\cite{Mariolis75,Davis96,DG97,DYB01}, co-ownership
networks of companies~\cite{KW99}, and collaborations of
scientists~\cite{GI95,DG99,BM00,Newman01a,Newman01b} and movie
actors~\cite{WS98}.  As well as having strongly non-Poisson degree
distributions~\cite{ASBS00,Newman01a}, these networks have a bipartite
structure; there are two distinct kinds of vertices on the graph with links
running only between vertices of unlike kinds~\cite{WF94}---see
Fig.~\ref{bifig}.  In the case of movie actors, for example, the two types
of vertices are movies and actors, and the network can be represented as a
graph with edges running between each movie and the actors that appear in
it.  Researchers have also considered the projection of this graph onto the
unipartite space of actors only, also called a one-mode
network~\cite{WF94}.  In such a projection two actors are considered
connected if they have appeared in a movie together.  The construction of
the one-mode network however involves discarding some of the information
contained in the original bipartite network, and for this reason it is more
desirable to model collaboration networks using the full bipartite
structure.

Given the high current level of interest in the structure of many of the
graphs described here, and given their substantial differences from the
ordinary random graphs that have been studied in the past, it would clearly
be useful if we could generalize the mathematics of random graphs to
non-Poisson degree distributions, and to directed and bipartite graphs.  In
this paper we do just that, demonstrating in detail how the statistical
properties of each of these graph types can be calculated exactly in the
limit of large graph size.  We also give examples of the application of our
theory to the modeling of a number of real-world networks, including the
world-wide web and collaboration graphs.

\section{Random graphs with arbitrary degree distributions}
\label{rg}
In this section we develop a formalism for calculating a variety of
quantities, both local and global, on large unipartite undirected graphs
with arbitrary probability distribution of the degrees of their vertices.
In all respects other than their degree distribution, these graphs are
assumed to be entirely random.  This means that the degrees of all vertices
are independent identically-distributed random integers drawn from a
specified distribution.  For a given choice of these degrees, also called
the ``degree sequence,'' the graph is chosen uniformly at random from the
set of all graphs with that degree sequence.  All properties calculated in
this paper are averaged over the ensemble of graphs generated in this way.
In the limit of large graph size an equivalent procedure is to study only
one particular degree sequence, averaging uniformly over all graphs with
that sequence, where the sequence is chosen to approximate as closely as
possible the desired probability distribution.  The latter procedure can be
thought of as a ``microcanonical ensemble'' for random graphs, where the
former is a ``canonical ensemble.''

Some results are already known for random graphs with arbitrary degree
distributions: in two beautiful recent papers~\cite{MR95,MR98}, Molloy and
Reed have derived formulas for the position of the phase transition at
which a giant component first appears, and the size of the giant component.
(These results are calculated within the microcanonical ensemble, but apply
equally to the canonical one in the large system size limit.)  The
formalism we present in this paper yields an alternative derivation of
these results and also provides a framework for obtaining other quantities
of interest, some of which we calculate.  In Sections~\ref{directed}
and~\ref{bipartite} we extend our formalism to the case of directed graphs
(such as the world-wide web) and bipartite graphs (such as collaboration
graphs).

\subsection{Generating functions}
\label{genfns}
Our approach is based on generating functions\cite{Wilf94}, the most
fundamental of which, for our purposes, is the generating function $G_0(x)$
for the probability distribution of vertex degrees~$k$.  Suppose that we
have a unipartite undirected graph---an acquaintance network, for
example---of $N$ vertices, with $N$ large.  We define
\begin{equation}
G_0(x) = \sum_{k=0}^\infty p_k x^k,
\label{defsg0}
\end{equation}
where $p_k$ is the probability that a randomly chosen vertex on the graph
has degree~$k$.  The distribution $p_k$ is assumed correctly normalized, so
that
\begin{equation}
G_0(1) = 1.
\end{equation}
The same will be true of all generating functions considered here, with a
few important exceptions, which we will note at the appropriate point.
Because the probability distribution is normalized and positive definite,
$G_0(x)$ is also absolutely convergent for all $|x|\le1$, and hence has no
singularities in this region.
All the calculations of this paper will be confined to the region
$|x|\le1$.

The function $G_0(x)$, and indeed any probability generating function, has
a number of properties that will prove useful in subsequent developments.

\medbreak\noindent {\bf Derivatives}\quad The probability $p_k$ is given by
the $k^{\rm th}$ derivative of $G_0$ according to
\begin{equation}
p_k = {1\over k!} {\d^k G_0\over\d x^k}\bigg|_{x=0}.
\label{derivatives}
\end{equation}
Thus the one function $G_0(x)$ encapsulates all the information contained
in the discrete probability distribution~$p_k$.  We say that the function
$G_0(x)$ ``generates'' the probability distribution~$p_k$.


\medbreak\noindent {\bf Moments}\quad The average over the probability
distribution generated by a generating function---for instance, the average
degree $z$ of a vertex in the case of $G_0(x)$---is given by
\begin{equation}
z = \av{k} = \sum_k k p_k = G_0'(1).
\label{avk}
\end{equation}
Thus if we can calculate a generating function we can also calculate the
mean of the probability distribution which it generates.  Higher moments of
the distribution can be calculated from higher derivatives also.  In
general, we have
\begin{equation}
\av{k^n} = \sum_k k^n p_k = 
  \biggl[ \biggl(x {\d\over\d x}\biggr)^{\!n} G_0(x) \biggr]_{x=1}.
\end{equation}

\medbreak\noindent {\bf Powers}\quad If the distribution of a property $k$
of an object is generated by a given generating function, then the
distribution of the total of $k$ summed over $m$ independent realizations
of the object is generated by the $m^{\rm th}$ power of that generating
function.  For example, if we choose $m$ vertices at random from a large
graph, then the distribution of the sum of the degrees of those vertices is
generated by~$[G_0(x)]^m$.  To see why this is so, consider the simple case
of just two vertices.  The square $[G_0(x)]^2$ of the generating function
for a single vertex can be expanded as
\begin{eqnarray}
[G_0(x)]^2 &=& \biggl[ \sum_k p_k x^k \biggr]^2
            =  \sum_{jk} p_j p_k x^{j+k}\nonumber\\
           &=& p_0 p_0 x^0 + (p_0 p_1 + p_1 p_0) x^1\nonumber\\
           & & + (p_0 p_2 + p_1 p_1 + p_2 p_0) x^2\nonumber\\
           & & + (p_0 p_3 + p_1 p_2 + p_2 p_1 + p_3 p_0) x^3
               + \ldots\nonumber\\
\end{eqnarray}
It is clear that the coefficient of the power of $x^n$ in this expression
is precisely the sum of all products $p_j p_k$ such that $j+k=n$, and hence
correctly gives the probability that the sum of the degrees of the two
vertices will be~$n$.  It is straightforward to convince oneself that this
property extends also to all higher powers of the generating function.

\medbreak All of these properties will be used in the derivations given in
this paper.

Another quantity that will be important to us is the distribution of the
degree of the vertices that we arrive at by following a randomly chosen
edge.  Such an edge arrives at a vertex with probability proportional to
the degree of that vertex, and the vertex therefore has a probability
distribution of degree proportional to $k p_k$.  The correctly normalized
distribution is generated by
\begin{equation}
{\sum_k k p_k x^k\over\sum_k k p_k} = x {G_0'(x)\over G_0'(1)}.
\label{arrival}
\end{equation}

If we start at a randomly chosen vertex and follow each of the edges at
that vertex to reach the $k$ nearest neighbors, then the vertices arrived
at each have the distribution of remaining outgoing edges generated by this
function, less one power of $x$, to allow for the edge that we arrived
along.  Thus the distribution of outgoing edges is generated by the
function
\begin{equation}
G_1(x) = {G_0'(x)\over G_0'(1)} = {1\over z}\, G_0'(x),
\label{defsg1}
\end{equation}
where $z$ is the average vertex degree, as before.  The probability that
any of these outgoing edges connects to the original vertex that we started
at, or to any of its other immediate neighbors, goes as $N^{-1}$ and hence
can be neglected in the limit of large~$N$.  Thus, making use of the
``powers'' property of the generating function described above, the
generating function for the probability distribution of the number of {\em
  second} neighbors of the original vertex can be written as
\begin{equation}
\sum_k p_k [G_1(x)]^k = G_0(G_1(x)).
\end{equation}
Similarly, the distribution of third-nearest neighbors is generated by
$G_0(G_1(G_1(x)))$, and so on.  The average number $z_2$ of second
neighbors is
\begin{equation}
z_2 = \biggl[ {\d\over\d x} G_0(G_1(x)) \biggr]_{x=1} = G_0'(1) G_1'(1)
    = G_0''(1),
\label{avk2}
\end{equation}
where we have made use of the fact that $G_1(1)=1$.  (One might be tempted
to conjecture that since the average number of first neighbors is
$G_0'(1)$, Eq.~\eref{avk}, and the average number of second neighbors is
$G_0''(1)$, Eq.~\eref{avk2}, then the average number of $m$th neighbors
should be given by the $m$th derivative of $G_0$ evaluated at $x=1$.  As we
show in Section~\ref{pathlength}, however, this conjecture is wrong.)

\subsection{Examples}
\label{examples}
To make things more concrete, we immediately introduce some examples of
specific graphs to illustrate how these calculations are carried out.

\paragraph{Poisson-distributed graphs} The simplest example of a graph of
this type is one for which the distribution of degree is binomial, or
Poisson in the large $N$ limit.  This distribution yields the standard
random graph studied by many mathematicians and discussed in
Section~\ref{intro}.  In this graph the probability $p=z/N$ of the
existence of an edge between any two vertices is the same for all vertices,
and $G_0(x)$ is given by
\begin{eqnarray}
G_0(x) &=& \sum_{k=0}^N \left({N\atop k}\right) p^k (1-p)^{N-k} x^k\nonumber\\
       &=& (1-p+px)^N = \e^{z(x-1)},
\label{poissongen}
\end{eqnarray}
where the last equality applies in the limit $N\to\infty$.  It is then
trivial to show that the average degree of a vertex is indeed $G_0'(1)=z$
and that the probability distribution of degree is given by $p_k =
z^k\e^{-z}/k!$, which is the ordinary Poisson distribution.  Notice also
that for this special case we have $G_1(x) = G_0(x)$, so that the
distribution of outgoing edges at a vertex is the same, regardless of
whether we arrived there by choosing a vertex at random, or by following a
randomly chosen edge.  This property, which is peculiar to the
Poisson-distributed random graph, is the reason why the theory of random
graphs of this type is especially simple.

\paragraph{Exponentially distributed graphs} Perhaps the next simplest type
of graph is one with an exponential distribution of vertex degrees
\begin{equation}
p_k = (1 - \e^{-1/\kappa}) \e^{-k/\kappa},
\label{expdist}
\end{equation}
where $\kappa$ is a constant.  The generating function for this
distribution is
\begin{equation}
G_0(x) = (1 - \e^{-1/\kappa}) \sum_{k=0}^\infty \e^{-k/\kappa} x^k
       = {1 - \e^{-1/\kappa}\over 1 - x\e^{-1/\kappa}},
\end{equation}
and
\begin{equation}
G_1(x) = \biggl[ {1 - \e^{-1/\kappa}\over 1 - x\e^{-1/\kappa}} \biggr]^2.
\end{equation}
An example of a graph with an exponential degree distribution is given in
Section~\ref{collabs}.

\paragraph{Power-law distributed graphs} The recent interest in the
properties of the world-wide web and of social networks leads us to
investigate the properties of graphs with a power-law distribution of
vertex degrees.  Such graphs have been discussed previously by
Barab\'asi~\etal~\cite{BA99,AJB99} and by Aiello~\etal\cite{ACL00}.  In this
paper, we will look at graphs with degree distribution given by
\begin{equation}
p_k = C k^{-\tau} \e^{-k/\kappa} \qquad \mbox{for $k\ge1$.}
\end{equation}
where $C$, $\tau$, and $\kappa$ are constants.  The reason for including
the exponential cutoff is two-fold: first many real-world graphs appear to
show this cutoff~\cite{ASBS00,Newman01a}; second it makes the distribution
normalizable for all $\tau$, and not just $\tau\ge2$.

The constant $C$ is fixed by the requirement of normalization, which gives
$C = [\Li_\tau(\e^{-1/\kappa})]^{-1}$ and hence
\begin{equation}
p_k = {k^{-\tau} \e^{-k/\kappa}\over\Li_\tau(\e^{-1/\kappa})} \qquad
\mbox{for $k\ge1$,}
\label{powerlaw}
\end{equation}
where $\Li_n(x)$ is the $n$th polylogarithm of $x$, a function familiar to
those who have worked with Feynman integrals.

Substituting~\eref{powerlaw} into Eq.~\eref{defsg0}, we find that the
generating function for graphs with this degree distribution is
\begin{equation}
G_0(x) = {\Li_\tau(x\e^{-1/\kappa})\over\Li_\tau(\e^{-1/\kappa})}.
\end{equation}
In the limit $\kappa\to\infty$---the case considered in
Refs.~\onlinecite{ACL00} and~\onlinecite{AJB99}---this simplifies to
\begin{equation}
G_0(x) = {\Li_\tau(x)\over\zeta(\tau)},
\end{equation}
where $\zeta(x)$ is the Riemann $\zeta$-function.

The function $G_1(x)$ is given by
\begin{equation}
G_1(x) = {\Li_{\tau-1}(x\e^{-1/\kappa})\over x\Li_{\tau-1}(\e^{-1/\kappa})}.
\end{equation}
Thus, for example, the average number of neighbors of a randomly-chosen
vertex is
\begin{equation}
z = G_0'(1) = {\Li_{\tau-1}(\e^{-1/\kappa})\over\Li_{\tau}(\e^{-1/\kappa})},
\label{plz1}
\end{equation}
and the average number of second neighbors is
\begin{equation}
z_2 = G_0''(1) = {\Li_{\tau-2}(\e^{-1/\kappa})-\Li_{\tau-1}(\e^{-1/\kappa})
                 \over\Li_{\tau}(\e^{-1/\kappa})}.
\label{plz2}
\end{equation}

\paragraph{Graphs with arbitrary specified degree distribution}
In some cases we wish to model specific real-world graphs which have known
degree distributions---known because we can measure them directly.  A
number of the graphs described in the introduction fall into this category.
For these graphs, we know the exact numbers $n_k$ of vertices having
degree~$k$, and hence we can write down the exact generating function for
that probability distribution in the form of a finite polynomial:
\begin{equation}
G_0(x) = {\sum_k n_k x^k\over\sum_k n_k},
\label{arbdist}
\end{equation}
where the sum in the denominator ensures that the generating function is
properly normalized.  As a example, suppose that in a community of 1000
people, each person knows between zero and five of the others, the exact
numbers of people in each category being, from zero to five: $\lbrace 86,
150, 363, 238, 109, 54 \rbrace$.  This distribution will then be generated
by the polynomial
\begin{equation}
G_0(x) = {86 + 150 x + 363 x^2 + 238 x^3 + 109x^4 + 54 x^5\over1000}.
\end{equation}

\begin{figure}
\begin{center}
\psfig{figure=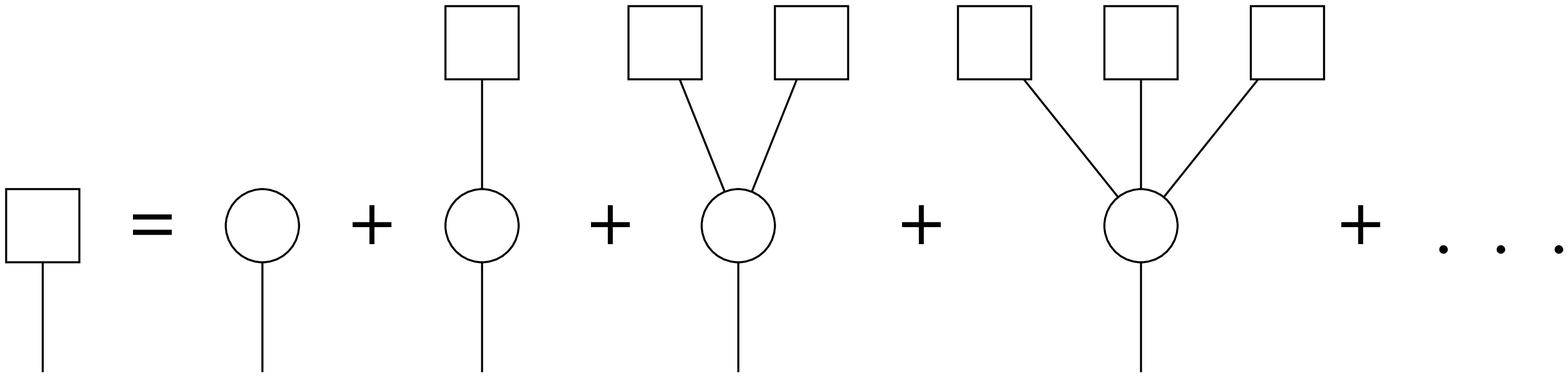,width=\columnwidth}
\end{center}
\caption{Schematic representation of the sum rule for the connected component
  of vertices reached by following a randomly chosen edge.  The probability
  of each such component (left-hand side) can be represented as the sum of
  the probabilities (right-hand side) of having only a single vertex,
  having a single vertex connected to one other component, or two other
  components, and so forth.  The entire sum can be expressed in closed form
  as Eq.~\eref{defsh1}.}
\label{sum}
\end{figure}

\subsection{Component sizes}
\label{compsizes}
We are now in a position to calculate some properties of interest for our
graphs.  First let us consider the distribution of the sizes of connected
components in the graph.  Let $H_1(x)$ be the generating function for the
distribution of the sizes of components which are reached by choosing a
random {\em edge\/} and following it to one of its ends.  We explicitly
exclude from $H_1(x)$ the giant component, if there is one; the giant
component is dealt with separately below.  Thus, except when we are
precisely at the phase transition where the giant component appears,
typical component sizes are finite, and the chances of a component
containing a closed loop of edges goes as $N^{-1}$, which is negligible in
the limit of large~$N$.  This means that the distribution of components
generated by $H_1(x)$ can be represented graphically as in Fig.~\ref{sum};
each component is tree-like in structure, consisting of the single site we
reach by following our initial edge, plus any number (including zero) of
other tree-like clusters, with the same size distribution, joined to it by
single edges.  If we denote by $q_k$ the probability that the initial site
has $k$ edges coming out of it other than the edge we came in along, then,
making use of the ``powers'' property of Section~\ref{genfns}, $H_1(x)$
must satisfy a self-consistency condition of the form
\begin{equation}
H_1(x) = x q_0 + x q_1 H_1(x) + x q_2 [H_1(x)]^2 + \ldots
\label{dyson}
\end{equation}
However, $q_k$ is nothing other than the coefficient of $x^k$ in the
generating function $G_1(x)$, Eq.~\eref{defsg1}, and hence Eq.~\eref{dyson}
can also be written
\begin{equation}
H_1(x) = x G_1(H_1(x)).
\label{defsh1}
\end{equation}

If we start at a randomly chosen {\em vertex,} then we have one such
component at the end of each edge leaving that vertex, and hence the
generating function for the size of the whole component is
\begin{equation}
H_0(x) = x G_0(H_1(x)).
\label{defsh0}
\end{equation}

In principle, therefore, given the functions $G_0(x)$ and $G_1(x)$, we can
solve Eq.~\eref{defsh1} for $H_1(x)$ and substitute into Eq.~\eref{defsh0}
to get $H_0(x)$.  Then we can find the probability that a randomly chosen
vertex belongs to a component of size $s$ by taking the $s$th derivative of
$H_0$.  In practice, unfortunately, this is usually impossible;
Eq.~\eref{defsh1} is a complicated and frequently transcendental equation,
which rarely has a known solution.  On the other hand, we note that the
coefficient of $x^s$ in the Taylor expansion of $H_1(x)$ (and therefore
also the $s$th derivative) are given exactly by only $s+1$ iterations of
Eq.~\eref{defsh0}, starting with $H_1=1$, so that the distribution
generated by $H_0(x)$ can be calculated exactly to finite order in finite
time.  With current symbolic manipulation programs, it is quite possible to
evaluate the first one hundred or so derivatives in this way.  Failing
this, an approximate solution can be found by numerical iteration and the
distribution of cluster sizes calculated from Eq.~\eref{derivatives} by
numerical differentiation.  Since direct evaluation of numerical
derivatives is prone to machine-precision problems, we recommend evaluating
the derivatives by numerical integration of the Cauchy formula, giving the
probability distribution $P_s$ of cluster sizes thus:
\begin{equation}
P_s = {1\over s!}\,{\d^s\!H_0\over\d z^s}\bigg|_{z=0}
     = {1\over2\pi\i} \oint {H_0(z)\over z^{s+1}} \d z.
\label{cauchy}
\end{equation}
The best numerical precision is obtained by using the largest possible
contour, subject to the condition that it enclose no poles of the
generating function.  The largest contour for which this condition is
satisfied in general is the unit circle $|z|=1$ (see Section~\ref{genfns}),
and we recommend using this contour for Eq.~\eref{cauchy}.  It is possible
to find the first thousand derivatives of a function without difficulty
using this method~\cite{MN00}.

\subsection{The mean component size, the phase transition, and the giant
  component}
\label{gcsec}
Although it is not usually possible to find a closed-form expression for
the complete distribution of cluster sizes on a graph, we can find
closed-form expressions for the {\em average\/} properties of clusters from
Eqs.~\eref{defsh1} and~\eref{defsh0}.  For example, the average size of the
component to which a randomly chosen vertex belongs, for the case where
there is no giant component in the graph, is given in the normal fashion by
\begin{equation}
\av{s} = H_0'(1) = 1 + G_0'(1) H_1'(1).
\end{equation}
From Eq.~\eref{defsh1} we have
\begin{equation}
H_1'(1) = 1 + G_1'(1) H_1'(1),
\end{equation}
and hence
\begin{equation}
\av{s} = 1 + {G_0'(1)\over1-G_1'(1)} = 1 + {z_1^2\over z_1-z_2},
\label{avs}
\end{equation}
where $z_1=z$ is the average number of neighbors of a vertex and $z_2$ is
the average number of second neighbors.  We see that this expression
diverges when
\begin{equation}
G_1'(1) = 1.
\label{perctrans}
\end{equation}
This point marks the phase transition at which a giant component first
appears.  Substituting Eqs.~\eref{defsg0} and~\eref{defsg1} into
Eq.~\eref{perctrans}, we can also write the condition for the phase
transition as
\begin{equation}
\sum_k k(k-2) p_k = 0.
\label{mrresult1}
\end{equation}
Indeed, since this sum increases monotonically as edges are added to the
graph, it follows that the giant component exists if and only if this sum
is positive.  This result has been derived by different means by Molloy and
Reed~\cite{MR95}.  An equivalent and intuitively reasonable statement,
which can also be derived from Eq.~\eref{avs}, is that the giant component
exists if and only if $z_2>z_1$.

Our generating function formalism still works when there is a giant
component in the graph, but, by definition, $H_0(x)$ then generates the
probability distribution of the sizes of components {\em excluding\/} the
giant component.  This means that $H_0(1)$ is no longer unity, as it is for
the other generating functions considered so far, but instead takes the
value $1-S$, where $S$ is the fraction of the graph occupied by the giant
component.  We can use this to calculate the size of the giant component
from Eqs.~\eref{defsh1} and~\eref{defsh0} thus:
\begin{equation}
S = 1 - G_0(u),
\label{giantcomp1}
\end{equation}
where $u\equiv H_1(1)$ is the smallest non-negative real solution of
\begin{equation}
u = G_1(u).
\label{giantcomp2}
\end{equation}
This result has been derived in a different but equivalent form by Molloy
and Reed~\cite{MR98}, using different methods.

The correct general expression for the average component size, excluding
the (formally infinite) giant component, if there is one, is
\begin{eqnarray}
\av{s} &=& {H_0'(1)\over H_0(1)}\nonumber\\
       &=& {1\over H_0(1)} \biggl[ G_0(H_1(1)) +
           {G_0'(H_1(1)) G_1(H_1(1))\over1 - G_1'(H_1(1))} \biggr]\nonumber\\
       &=& 1 + {zu^2\over [1-S][1-G_1'(u)]},
\end{eqnarray}
which is equivalent to~\eref{avs} when there is no giant component ($S=0$,
$u=1$).

For example, in the ordinary random graph with Poisson degree distribution,
we have $G_0(x) = G_1(x) = \e^{z(x-1)}$ (Eq.~\eref{poissongen}), and hence
we find simply that $1-S=u$ is a solution of $u=G_0(u)$, or equivalently
that
\begin{equation}
S = 1 - \e^{-zS}.
\end{equation}
The average component size is given by
\begin{equation}
\av{s} = {1\over1-z+zS}.
\end{equation}
These are all well-known results~\cite{Bollobas85}.

For graphs with purely power-law distributions (Eq.~\eref{powerlaw} with
$\kappa\to\infty$), $S$~is given by~\eref{giantcomp1} with $u$ the smallest
non-negative real solution of
\begin{equation}
u = {\Li_{\tau-1}(u)\over u\zeta(\tau-1)}.
\end{equation}
For all $\tau\le2$ this gives $u=0$, and hence $S=1$, implying that a
randomly chosen vertex belongs to the giant component with probability
tending to~1 as $\kappa\to\infty$.  For graphs with $\tau>2$, the
probability of belonging to the giant component is strictly less than~1,
even for infinite~$\kappa$.  In other words, the giant component
essentially fills the entire graph for $\tau\le2$, but not for $\tau>2$.
These results have been derived by different means by
Aiello~\etal~\cite{ACL00}.

\subsection{Asymptotic form of the cluster size distribution}
\label{asymptotic}
A variety of results are known about the asymptotic properties of the
coefficients of generating functions, some of which can usefully be applied
to the distribution of cluster sizes $P_s$ generated by $H_0(x)$.  Close to
the phase transition, we expect the tail of the distribution $P_s$ to
behave as
\begin{equation}
P_s \sim s^{-\alpha} \e^{-s/s^*},
\label{tail}
\end{equation}
where the constants $\alpha$ and $s^*$ can be calculated from the
properties of $H_0(x)$ as follows.

The cutoff parameter $s^*$ is simply related to the radius of
convergence~$|x^*|$ of the generating function~\cite{Wilf94,HL14}, according
to
\begin{equation}
s^* = {1\over\log |x^*|}.
\label{rofc}
\end{equation}
The radius of convergence $|x^*|$ is equal to the magnitude of the position
$x^*$ of the singularity in $H_0(x)$ nearest to the origin.  From
Eq.~\eref{defsh0} we see that such a singularity may arise either through a
singularity in $G_0(x)$ or through one in $H_1(x)$.  However, since the
first singularity in $G_0(x)$ is known to be outside the unit circle
(Section~\ref{genfns}), and the first singularity in $H_1(x)$ tends to
$x=1$ as we go to the phase transition (see below), it follows that,
sufficiently close to the phase transition, the singularity in $H_0(x)$
closest to the origin is also a singularity in $H_1(x)$.  With this result
$x^*$ is easily calculated.

Although we do not in general have a closed-form expression for $H_1(x)$,
it is easy to derive one for its functional inverse.  Putting $w=H_1(x)$
and $x=H_1^{-1}(w)$ in Eq.~\eref{defsh1} and rearranging, we find
\begin{equation}
x = H_1^{-1}(w) = {w\over G_1(w)}.
\label{inverse}
\end{equation}
The singularity of interest corresponds to the point $w^*$ at which the
derivative of $H_1^{-1}(w)$ is zero, which is a solution of
\begin{equation}
G_1(w^*) - w^* G_1'(w^*) = 0.
\label{singularity}
\end{equation}
Then $x^*$ (and hence $s^*$) is given by Eq.~\eref{inverse}.  Note that
there is no guarantee that~\eref{singularity} has a finite solution, and
that if it does not, then $P_s$ will not in general follow the form of
Eq.~\eref{tail}.

When we are precisely at the phase transition of our system, we have
$G_1(1)=G_1'(1)=1$, and hence the solution of Eq.~\eref{singularity} gives
$w^*=x^*=1$---a result which we used above---and $s^*\to\infty$.  We can
use the fact that $x^*=1$ at the transition to calculate the value of the
exponent $\alpha$ as follows.  Expanding $H_1^{-1}(w)$ about $w^*=1$ by
putting $w=1+\epsilon$ in Eq.~\eref{inverse}, we find that
\begin{equation}
H_1^{-1}(1+\epsilon) = 1 - \half G_1''(1) \epsilon^2 + \O(\epsilon^3),
\label{expansion}
\end{equation}
where we have made use of $G_1(1)=G_1'(1)=1$ at the phase transition.  So
long as $G_1''(1)\ne0$, which in general it is not, this implies that
$H_1(x)$ and hence also $H_0(x)$ are of the form
\begin{equation}
H_0(x) \sim (1-x)^\beta \qquad \mbox{as $x\to1$,}
\end{equation}
with $\beta=\frac12$.  This exponent is related to the exponent $\alpha$ as
follows.  Equation~\eref{tail} implies that $H_0(x)$ can be written in the
form
\begin{equation}
H_0(x) = \sum_{s=0}^{a-1} P_s x^s + C \sum_{s=a}^\infty s^{-\alpha}
       \e^{-s/s^*} x^s + \epsilon(a),
\label{parts}
\end{equation}
where $C$ is a constant and the last (error) term $\epsilon(a)$ is assumed
much smaller than the second term.  The first term in this expression is a
finite polynomial and therefore has no singularities on the finite plane;
the singularity resides in the second term.  Using this equation, the
exponent $\beta$ can be written:
\begin{eqnarray}
\beta &=& \lim_{x\to1} \biggl[ 1 + (x-1) {H_0''(x)\over H_0'(x)}
                        \biggr]\nonumber\\
      &=& \lim_{a\to\infty}\lim_{x\to1} \biggl[ {1\over x} +
           {x-1\over x} {\sum_{s=a}^\infty s^{2-\alpha} x^{s-1}\over
           \sum_{s=a}^\infty s^{1-\alpha} x^{s-1}} \biggr]\nonumber\\
      &=&  \lim_{a\to\infty}\lim_{x\to1} \biggl[ {1\over x} +
           {1-x\over x\log x} {\Gamma(3-\alpha,-a\log x)\over
                                  \Gamma(2-\alpha,-a\log x)}\biggr],
\end{eqnarray}
where we have replaced the sums with integrals as $a$ becomes large, and
$\Gamma(\nu,\mu)$ is the incomplete $\Gamma$-function.  Taking the limits
in the order specified and rearranging for $\alpha$, we then get
\begin{equation}
\alpha = \beta + 1 = \mbox{$\frac32$},
\label{valuealpha}
\end{equation}
regardless of degree distribution, except in the special case where
$G_1''(1)$ vanishes (see Eq.~\eref{expansion}).  The result
$\alpha=\frac32$ was known previously for the ordinary Poisson random
graph~\cite{Bollobas85}, but not for other degree distributions.

\subsection{Numbers of neighbors and average path length}
\label{pathlength}
We turn now to the calculation of the number of neighbors who are $m$ steps
away from a randomly chosen vertex.  As shown in Section~\ref{genfns}, the
probability distributions for first- and second-nearest neighbors are
generated by the functions $G_0(x)$ and $G_0(G_1(x))$.  By extension, the
distribution of $m$th neighbors is generated by $G_0(G_1(\ldots
G_1(x)\ldots))$, with $m-1$ iterations of the function $G_1$ acting on
itself.  If we define $G^{(m)}(x)$ to be this generating function for $m$th
neighbors, then we have
\begin{equation}
G^{(m)}(x) = \biggl\lbrace \begin{array}{ll}
               G_0(x)                   & \mbox{for $m=1$}\\
               G^{(m-1)}(G_1(x)) \qquad & \mbox{for $m\ge2$.}
             \end{array}
\end{equation}
Then the average number $z_m$ of $m$th-nearest neighbors is
\begin{equation}
z_m = {\d G^{(m)}\over\d x}\bigg|_{x=1} = G_1'(1) {G^{(m-1)}}'(1)
    = G_1'(1) z_{m-1}.
\end{equation}
Along with the initial condition $z_1=z=G_0'(1)$, this then tells us that
\begin{equation}
z_m = [G_1'(1)]^{m-1} G_0'(1) = \biggl[{z_2\over z_1}\biggr]^{m-1} z_1.
\label{zm}
\end{equation}
From this result we can make an estimate of the typical length $\ell$ of
the shortest path between two randomly chosen vertices on the graph.  This
typical path length is reached approximately when the total number of
neighbors of a vertex out to that distance is equal to the number of
vertices on the graph, i.e.,~when
\begin{equation}
1 + \sum_{m=1}^\ell z_m = N.
\label{sumell}
\end{equation}
Using Eq.~\eref{zm} this gives us
\begin{equation}
\ell = {\log [(N-1)(z_2-z_1)+z_1^2] - \log z_1^2\over\log (z_2/z_1)}.
\label{fullell}
\end{equation}
In the common case where $N\gg z_1$ and $z_2\gg z_1$, this reduces to
\begin{equation}
\ell = {\log (N/z_1)\over\log (z_2/z_1)} + 1.
\label{ell}
\end{equation}
This result is only approximate for two reasons.  First, the conditions
used to derive it are only an approximation; the exact answer depends on
the detailed structure of the graph.  Second, it assumes that all vertices
are reachable from a randomly chosen starting vertex.  In general however
this will not be true.  For graphs with no giant component it is certainly
not true and Eq.~\eref{ell} is meaningless.  Even when there is a giant
component however, it is usually not the case that it fills the entire
graph.  A better approximation to $\ell$ may therefore be given by
replacing $N$ in Eq.~\eref{ell} by $NS$, where $S$ is the fraction of the
graph occupied by the giant component, as in Section~\ref{gcsec}.

Such shortcomings notwithstanding, there are a number of remarkable
features of Eq.~\eref{ell}:
\begin{enumerate}
\item It shows that the average vertex--vertex distance for all random
  graphs, regardless of degree distribution, should scale logarithmically
  with size $N$, according to $\ell = A + B\log N$, where $A$ and $B$ are
  constants.  This result is of course well-known for a number of special
  cases.
\item It shows that the average distance, which is a global property, can
  be calculated from a knowledge only of the average numbers of first- and
  second-nearest neighbors, which are local properties.  It would be
  possible therefore to measure these numbers empirically by purely local
  measurements on a graph such as an acquaintance network and from them to
  determine the expected average distance between vertices.  For some
  networks at least, this gives a surprisingly good estimate of the true
  average distance~\cite{Newman01b}.
\item It shows that only the average numbers of first- and second-nearest
  neighbors are important to the calculation of average distances, and thus
  that two random graphs with completely different distributions of vertex
  degrees, but the same values of $z_1$ and $z_2$, will have the same
  average distances.
\end{enumerate}

For the case of the purely theoretical example graphs we discussed earlier,
we cannot make an empirical measurement of $z_1$ and $z_2$, but we can
still employ Eq.~\eref{ell} to calculate~$\ell$.  In the case of the
ordinary (Poisson) random graph, for instance, we find from
Eq.~\eref{poissongen} that $z_1=z$, $z_2=z^2$, and so $\ell=\log N/\log z$,
which is the standard result for graphs of this type~\cite{Bollobas85}.
For the graph with degree distributed according to the truncated power law,
Eq.~\eref{powerlaw}, $z_1$ and $z_2$ are given by Eqs.~\eref{plz1}
and~\eref{plz2}, and the average vertex--vertex distance is
\begin{equation}
\ell = {\log N +
        \log \bigl[{\Li_{\tau}(\e^{-1/\kappa})/
             \Li_{\tau-1}(\e^{-1/\kappa})}\bigr]
       \over \log \bigl[{\Li_{\tau-2}(\e^{-1/\kappa})/
                   \Li_{\tau-1}(\e^{-1/\kappa})}-1\bigr]} + 1.
\end{equation}
In the limit $\kappa\to\infty$, this becomes
\begin{equation}
\ell = {\log N + \log\bigl[\zeta(\tau)/\zeta(\tau-1)\bigr]\over
        \log\bigl[\zeta(\tau-2)/\zeta(\tau-1)-1\bigr]} + 1.
\end{equation}
Note that this expression does not have a finite positive real value for
any $\tau<3$, indicating that one must specify a finite cutoff $\kappa$ for
the degree distribution to get a well-defined average vertex--vertex
distance on such graphs.

\subsection{Simulation results}
\label{undirsim}
As a check on the results of this section, we have performed extensive
computer simulations of random graphs with various distributions of vertex
degree.  Such graphs are relatively straightforward to generate.  First, we
generate a set of $N$ random numbers $\set{k_i}$ to represent the degrees
of the $N$ vertices in the graph.  These may be thought of as the ``stubs''
of edges, emerging from their respective vertices.  Then we choose pairs of
these stubs at random and place edges on the graph joining them up.  It is
simple to see that this will generate all graphs with the given set of
vertex degrees with equal probability.  The only small catch is that the
sum $\sum_i k_i$ of the degrees must be even, since each edge added to the
graph must have two ends.  This is not difficult to contrive however.  If
the set $\set{k_i}$ is such that the sum is odd, we simply throw it away
and generate a new set.

As a practical matter, integers representing vertex degrees with any
desired probability distribution can be generated using the transformation
method if applicable, or failing that, a rejection or hybrid
method~\cite{NB99}.  For example, degrees obeying the power-law-plus-cutoff
form of Eq.~\eref{powerlaw} can be generated using a two-step hybrid
transformation/rejection method as follows.  First, we generate random
integers $k\ge1$ with distribution proportional to $\e^{-k/\kappa}$ using
the transformation\cite{note1}
\begin{equation}
k = \lceil -\kappa \log(1-r) \rceil,
\label{expdeviate}
\end{equation}
where $r$ is a random real number uniformly distributed in the range $0\le
r<1$.  Second, we accept this number with probability $k^{-\tau}$, where by
``accept'' we mean that if the number is not accepted we discard it and
generate another one according to Eq.~\eref{expdeviate}, repeating the
process until one is accepted.

\begin{figure}
\begin{center}
\psfig{figure=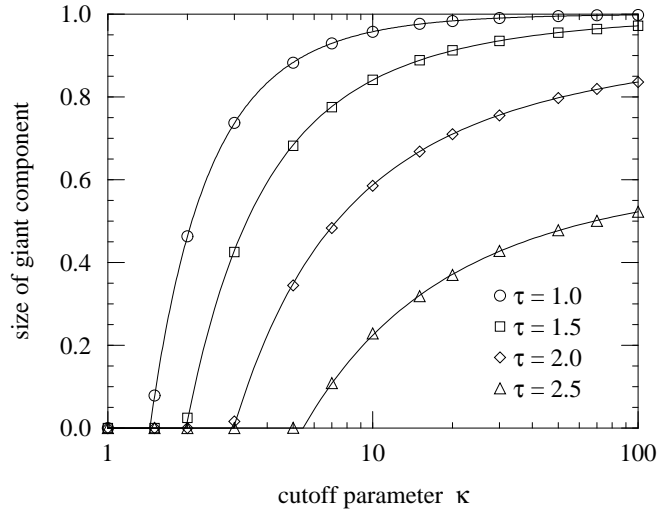,width=\columnwidth}
\end{center}
\caption{The size of the giant component in random graphs with vertex
  degrees distributed according to Eq.~\eref{powerlaw}, as a function of
  the cutoff parameter $\kappa$ for five different values of the exponent
  $\tau$.  The points are results from numerical simulations on graphs of
  $N=1\,000\,000$ vertices, and the solid lines are the theoretical value
  for infinite graphs, Eqs.~\eref{giantcomp1} and~\eref{giantcomp2}.  The
  error bars on the simulation results are smaller than the data points.}
\label{unipl}
\end{figure}

In Fig.~\ref{unipl} we show results for the size of the giant component in
simulations of undirected unipartite graphs with vertex degrees distributed
according to Eq.~\eref{powerlaw} for a variety of different values of
$\tau$ and~$\kappa$.  On the same plot we also show the expected value of
the same quantity derived by numerical solution of Eqs.~\eref{giantcomp1}
and~\eref{giantcomp2}.  As the figure shows, the agreement between
simulation and theory is excellent.

\section{Directed graphs}
\label{directed}
We turn now to directed graphs with arbitrary degree distributions.  An
example of a directed graph is the world-wide web, since every hyperlink
between two pages on the web goes in only one direction.  The web has a
degree distribution that follows a power-law, as discussed in
Section~\ref{intro}.

Directed graphs introduce a subtlety that is not present in undirected
ones, and which becomes important when we apply our generating function
formalism.  In a directed graph it is not possible to talk about a
``component''---i.e.,~a group of connected vertices---because even if
vertex~A can be reached by following (directed) edges from vertex~B, that
does not necessarily mean that vertex~B can be reached from vertex~A.
There are two correct generalizations of the idea of the component to a
directed graph: the set of vertices which are reachable from a given
vertex, and the set from which a given vertex can be reached.  We will
refer to these as ``out-components'' and ``in-components'' respectively.
An in-component can also be thought of as those vertices reachable by
following edges backwards (but not forwards) from a specified vertex.  It
is possible to study directed graphs by allowing both forward and backward
traversal of edges (see Ref.~\onlinecite{BKMRRSTW00}, for example).  In
this case, however, the graph effectively becomes undirected and should be
treated with the formalism of Section~\ref{rg}.

With these considerations in mind, we now develop the generating function
formalism appropriate to random directed graphs with arbitrary degree
distributions.

\subsection{Generating functions}
\label{dirgf}
In a directed graph, each vertex has separate in-degree and out-degree for
links running into and out of that vertex.  Let us define $p_{jk}$ to be
the probability that a randomly chosen vertex has in-degree $j$ and
out-degree~$k$.  It is important to realize that in general this joint
distribution of $j$ and $k$ is not equal to the product $p_j p_k$ of the
separate distributions of in- and out-degree.  In the world-wide web, for
example, it seems likely (although this question has not been investigated
to our knowledge) that sites with a large number of outgoing links also
have a large number of incoming ones, i.e.,~that $j$ and $k$ are
correlated, so that $p_{jk}\ne p_j p_k$.  We appeal to those working on
studies of the structure of the web to measure the joint distribution of
in- and out-degrees of sites; empirical data on this distribution would
make theoretical work much easier!

We now define a generating function for the joint probability distribution
of in- and out-degrees, which is necessarily a function of two independent
variables, $x$ and $y$, thus:
\begin{equation}
\cG(x,y) = \sum_{jk} p_{jk} x^j y^k.
\label{jointgen}
\end{equation}
Since every edge on a directed graph must leave some vertex and enter
another, the net average number of edges entering a vertex is zero, and
hence $p_{jk}$ must satisfy the constraint
\begin{equation}
\sum_{jk} (j-k) p_{jk} = 0.
\label{inoutfix}
\end{equation}
This implies that $\cG(x,y)$ must satisfy
\begin{equation}
{\partial \cG\over\partial x}\bigg|_{x,y=1} =
{\partial \cG\over\partial y}\bigg|_{x,y=1} = z,
\label{sumrule}
\end{equation}
where $z$ is the average degree (both in and out) of vertices in the graph.

Using the function $\cG(x,y)$, we can, as before, define generating
functions $G_0$ and $G_1$ for the number of out-going edges leaving a
randomly chosen vertex, and the number leaving the vertex reached by
following a randomly chosen edge.  We can also define generating functions
$F_0$ and $F_1$ for the number arriving at such a vertex.  These functions
are given by
\begin{eqnarray}
\label{defsf0f1}
F_0(x) &=& \cG(x,1),\qquad
F_1(x) = 
{1\over z}\, {\partial \cG\over\partial y}\bigg|_{y=1},\\
\label{defsg0g1}
G_0(y) &=& \cG(1,y),\qquad
G_1(y) = 
{1\over z}\, {\partial \cG\over\partial x}\bigg|_{x=1}.
\end{eqnarray}
Once we have these functions, many results follow as before.  The average
numbers of first and second neighbors reachable from a randomly chosen
vertex are given by Eq.~\eref{sumrule} and
\begin{equation}
z_2 = G_0'(1) G_1'(1) 
    = {\partial^2\cG\over\partial x\partial y}\bigg|_{x,y=1}.
\label{dirz2}
\end{equation}
These are also the numbers of first and second neighbors from which a
random vertex can be reached, since Eqs.~\eref{sumrule} and~\eref{dirz2}
are manifestly symmetric in $x$ and $y$.  We can also make an estimate of
the average path length on the graph from
\begin{equation}
\ell = {\log (N/z_1)\over\log (z_2/z_1)} + 1,
\label{ell2}
\end{equation}
as before.  However, this equation should be used with caution.  As
discussed in Section~\ref{pathlength}, the derivation of this formula
assumes that we are in a regime in which the bulk of the graph is reachable
from most vertices.  On a directed graph however, this may be far from
true, as appears to be the case with the world-wide web~\cite{BKMRRSTW00}.

The probability distribution of the numbers of vertices reachable from a
randomly chosen vertex in a directed graph---i.e.,~of the sizes of the
out-components---is generated by the function $H_0(y) = y G_0(H_1(y))$,
where $H_1(y)$ is a solution of $H_1(y) = y G_1(H_1(y))$, just as before.
(A similar and obvious pair of equations governs the sizes of the
in-components.)  The results for the asymptotic behavior of the component
size distribution from Section~\ref{asymptotic} generalize
straightforwardly to directed graphs.  The average out-component size for
the case where there is no giant component is given by Eq.~\eref{avs}, and
thus the point at which a giant component first appears is given once more
by $G_1'(1) = 1$.  Substituting Eq.~\eref{jointgen} into this expression
gives the explicit condition
\begin{equation}
\sum_{jk} (2jk-j-k) p_{jk} = 0
\label{mrresult2}
\end{equation}
for the first appearance of the giant component.  This expression is the
equivalent for the directed graph of Eq.~\eref{mrresult1}.  It is also
possible, and equally valid, to define the position at which the giant
component appears by $F_1'(1)=1$, which provides an alternative derivation
for Eq.~\eref{mrresult2}.

\begin{figure}
\begin{center}
\psfig{figure=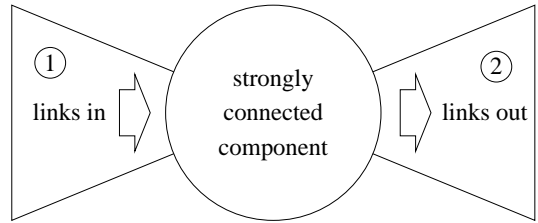,width=7cm}
\end{center}
\caption{The ``bow-tie'' diagram proposed by Broder~\etal\ as a
  representation of the giant component of the world-wide web (although it
  can be used to visualize any directed graph).}
\label{bowtie}
\end{figure}

Just as with the individual in- and out-components for vertices, the size
of the giant component on a directed graph can also be defined in different
ways.  The giant component can be represented using the ``bow-tie'' diagram
of Broder~\etal~\cite{BKMRRSTW00}, which we depict (in simplified form) in
Fig.~\ref{bowtie}.  The diagram has three parts.  The strongly connected
portion of the giant component, represented by the central circle, is that
portion in which every vertex can be reached from every other.  The two
sides of the bow-tie represent (1)~those vertices from which the strongly
connected component can be reached but which it is not possible to reach
from the strongly connected component and (2)~those vertices which can be
reached from the strongly connected component but from which it is not
possible to reach the strongly connected component.  The solution of
Eqs.~\eref{giantcomp1} and~\eref{giantcomp2} with $G_0(x)$ and $G_1(x)$
defined according to Eq.~\eref{defsg0g1} gives the number of vertices, as a
fraction of~$N$, in the giant strongly connected component plus those
vertices {\em from which\/} the giant strongly connected component can be
reached.  Using $F_0(x)$ and $F_1(x)$ (Eq.~\eref{defsf0f1}) in place of
$G_0(x)$ and $G_1(x)$ gives a different solution, which represents the
fraction of the graph in the giant strongly connected component plus those
vertices which can be reached from it.

\begin{figure}
\begin{center}
\psfig{figure=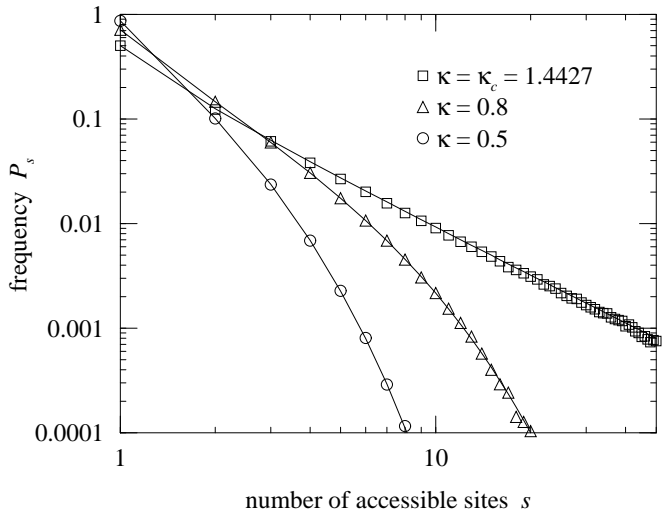,width=\columnwidth}
\end{center}
\caption{The distribution $P_s$ of the numbers of vertices accessible from
  each vertex of a directed graph with identically exponentially
  distributed in- and out-degree.  The points are simulation results for
  systems of $N=1\,000\,000$ vertices and the solid lines are the analytic
  solution.}
\label{direxp}
\end{figure}

\subsection{Simulation results}
\label{dirsim}
We have performed simulations of directed graphs as a check on the results
above.  Generation of random directed graphs with known joint degree
distribution $p_{jk}$ is somewhat more complicated than generation of
undirected graphs discussed in Section~\ref{undirsim}.  The method we use
is as follows.  First, it is important to ensure that the averages of the
distributions of in- and out-degree of the graph are the same, or
equivalently that $p_{jk}$ satisfies Eq.~\eref{inoutfix}.  If this is not
the case, at least to good approximation, then generation of the graph will
be impossible.  Next, we generate a set of $N$ in/out-degree pairs
$(j_i,k_i)$, one for each vertex~$i$, according to the joint distribution
$p_{jk}$, and calculate the sums $\sum_i j_i$ and $\sum_i k_i$.  These sums
are required to be equal if there are to be no dangling edges in the graph,
but in most cases we find that they are not.  To rectify this we use a
simple procedure.  We choose a vertex $i$ at random, discard the numbers
$(j_i,k_i)$ for that vertex and generate new ones from the distribution
$p_{jk}$.  We repeat this procedure until the two sums are found to be
equal.  Finally, we choose random in/out pairs of edges and join them
together to make a directed graph.  The resulting graph has the desired
number of vertices and the desired joint distribution of in and out degree.

We have simulated directed graphs in which the distribution $p_{jk}$ is
given by a simple product of independent distributions of in- and
out-degree.  (As pointed out in Section~\ref{dirgf}, this is not generally
the case for real-world directed graphs, where in- and out-degree may be
correlated.)  In Fig.~\ref{direxp} we show results from simulations of
graphs with identically distributed (but independent) in- and out-degrees
drawn from the exponential distribution, Eq.~\eref{expdist}.  For this
distribution, solution of the critical-point equation $G_1'(1)=1$ shows
that the giant component first appears at $\kappa_c=[\log 2]^{-1}=1.4427$.
The three curves in the figure show the distribution of numbers of vertices
accessible from each vertex in the graph for $\kappa=0.5$, $0.8$,
and~$\kappa_c$.  The critical distribution follows a power-law form (see
Section~\ref{compsizes}), while the others show an exponential cutoff.  We
also show the exact distribution derived from the coefficients in the
expansion of $H_1(x)$ about zero.  Once again, theory and simulation are in
good agreement.  A fit to the distribution for the case $\kappa=\kappa_c$
gives a value of $\alpha=1.50\pm0.02$, in good agreement with
Eq.~\eref{valuealpha}.


\section{Bipartite graphs}
\label{bipartite}
The collaboration graphs of scientists, company directors, and movie actors
discussed in Section~\ref{intro} are all examples of bipartite graphs.  In
this section we study the theory of bipartite graphs with arbitrary degree
distributions.  To be concrete, we will speak in the language of ``actors''
and ``movies,'' but clearly all the developments here are applicable to
academic collaborations, boards of directors, or any other bipartite graph
structure.

\subsection{Generating functions and basic results}
\label{bigenfns}
Consider then a bipartite graph of $M$ movies and $N$ actors, in which each
actor has appeared in an average of $\mu$ movies and each movie has a cast
of average size $\nu$ actors.  Note that only three of these parameters are
independent, since the fourth is given by the equality
\begin{equation}
{\mu\over M} = {\nu\over N}.
\label{mMnN}
\end{equation}
Let $p_j$ be the probability distribution of the degree of actors (i.e.,~of
the number of movies in which they have appeared) and $q_k$ be the
distribution of degree (i.e.,~cast size) of movies.  We define two
generating functions which generate these probability distributions thus:
\begin{equation}
f_0(x) = \sum_j p_j x^j,\qquad g_0(x) = \sum_k q_k x^k.
\label{defsf0g0}
\end{equation}
(It may be helpful to think of $f$ as standing for ``film,'' in order to
keep these two straight.)  As before, we necessarily have
\begin{equation}
f_0(1) = g_0(1) = 1,\quad f_0'(1) = \mu,\quad g_0'(1) = \nu.
\end{equation}

If we now choose a random edge on our bipartite graph and follow it both
ways to reach the movie and actor which it connects, then the distribution
of the number of other edges leaving those two vertices is generated by the
equivalent of~\eref{defsg1}:
\begin{equation}
f_1(x) = {1\over\mu}\,f_0'(x),\qquad g_1(x) = {1\over\nu}\,g_0'(x).
\label{defsf1g1}
\end{equation}
Now we can write the generating function for the distribution of the number
of co-stars (i.e.,~actors in shared movies) of a randomly chosen actor as
\begin{equation}
G_0(x) = f_0(g_1(x)).
\label{defscg0}
\end{equation}
If we choose a random edge, then the distribution of number of co-stars of
the actor to which it leads is generated by
\begin{equation}
G_1(x) = f_1(g_1(x)).
\end{equation}
These two functions play the same role in the one-mode network of actors as
the functions of the same name did for the unipartite random graphs of
Section~\ref{rg}.  Once we have calculated them, all the results from
Section~\ref{rg} follow exactly as before.

The numbers of first and second neighbors of a randomly chosen actor are
\begin{eqnarray}
\label{biavz}
z_1 &=& G_0'(1) = f_0'(1) g_1'(1),\\
z_2 &=& G_0'(1) G_1'(1) = f_0'(1) f_1'(1) [g_1'(1)]^2.
\end{eqnarray}
Explicit expressions for these quantities can be obtained by substituting
from Eqs.~\eref{defsf0g0} and~\eref{defsf1g1}.  The average vertex--vertex
distance on the one-mode graph is given as before by Eq.~\eref{ell}.  Thus,
it is possible to estimate average distances on such graphs by measuring
only the numbers of first and second neighbors.

The distribution of the sizes of the connected components in the one-mode
network is generated by Eq.~\eref{defsh0}, where $H_1(x)$ is a solution of
Eq.~\eref{defsh1}.  The asymptotic results of Section~\ref{asymptotic}
generalize simply to the bipartite case, and the average size of a
connected component in the absence of a giant component is
\begin{equation}
\av{s} = 1 + {G_0'(1)\over1-G_1'(1)},
\label{biavs}
\end{equation}
as before.  This diverges when $G_1'(1)=1$, marking the first appearance of
the giant component.  Equivalently, the giant component first appears when
\begin{equation}
f_0''(1) g_0''(1) = f_0'(1) g_0'(1).
\label{bitransition}
\end{equation}
Substituting from Eq.~\eref{defsf0g0}, we then derive the explicit
condition for the first appearance of the giant component:
\begin{equation}
\sum_{jk} jk(jk-j-k) p_j q_k = 0.
\end{equation}
The size $S$ of the giant component, as a fraction of the total number $N$
of actors, is given as before by the solution of Eqs.~\eref{giantcomp1}
and~\eref{giantcomp2}.

Of course, all of these results work equally well if ``actors'' and
``movies'' are interchanged.  One can calculate the average distance
between movies in terms of common actors shared, the size and distribution
of connected components of movies, and so forth, using the formulas given
above, with only the exchange of $f_0$ and $f_1$ for $g_0$ and $g_1$.  The
formula~\eref{bitransition} is, not surprisingly, invariant under this
interchange, so that the position of the onset of the giant component is
the same regardless of whether one is looking at actors or movies.

\subsection{Clustering}
Watts and Strogatz\cite{WS98} have introduced the concept of clustering in
social networks, also sometimes called network transitivity.  Clustering
refers to the increased propensity of pairs of people to be acquainted with
one another if they have another acquaintance in common.  Watts and
Strogatz defined a clustering coefficient which measures the degree of
clustering on a graph.  For our purposes, the definition of this
coefficient is
\begin{equation}
C = {\mbox{$3\times$ number of triangles on the graph}\over
     \mbox{number of connected triples of vertices}}
  = {3 N_\triangle\over N_3}.
\label{defsc}
\end{equation}
Here ``triangles'' are trios of vertices each of which is connected to both
of the others, and ``connected triples'' are trios in which at least one is
connected to both the others.  The factor of 3 in the numerator accounts
for the fact that each triangle contributes to three connected triples of
vertices, one for each of its three vertices.  With this factor of 3, the
value of $C$ lies strictly in the range from zero to one.  In the directed
and undirected unipartite random graphs of Sections~\ref{rg}
and~\ref{directed}, $C$ is trivially zero in the limit $N\to\infty$.  In
the one-mode projections of bipartite graphs, however, both the actors and
the movies can be expected to have non-zero clustering.  We here treat the
case for actors.  The case for movies is easily derived by swapping $f$s
and $g$s.

An actor who has $z\equiv z_1$ co-stars in total contributes $\frac12
z(z-1)$ connected triples to $N_3$, so that
\begin{equation}
N_3 = \half N \sum_z z(z-1) r_z,
\end{equation}
where $r_z$ is the probability of having $z$ co-stars.  As shown above
(Eq.~\eref{defscg0}), the distribution $r_z$ is generated by $G_0(x)$ and
so
\begin{equation}
N_3 = \half N G_0''(1).
\end{equation}

A movie which stars $k$ actors contributes $\frac16 k(k-1)(k-2)$ triangles
to the total triangle count in the one-mode graph.  Thus the total number
of triangles on the graph is the sum of $\frac16 k(k-1)(k-2)$ over all
movies, which is given by
\begin{equation}
N_\triangle = \mbox{$\frac16$} M \sum_k k(k-1)(k-2) q_k
            = \mbox{$\frac16$} M g_0'''(1).
\end{equation}

Substituting into Eq.~\eref{defsc}, we then get
\begin{equation}
C = {M\over N}\,{g_0'''(1)\over G_0''(1)}.
\label{bicc}
\end{equation}
Making use of Eqs.~\eref{mMnN}, \eref{defsf0g0}, and~\eref{defscg0}, this
can also be written as
\begin{equation}
{1\over C} - 1 =
{(\mu_2-\mu_1)(\nu_2-\nu_1)^2\over\mu_1\nu_1(2\nu_1-3\nu_2+\nu_3)},
\end{equation}
where $\mu_n=\sum_k k^n p_k$ is the $n$th moment of the distribution of
numbers of movies in which actors have appeared, and $\nu_n$ is the same
for cast size (number of actors in a movie).

\subsection{Example}
\label{biexample}
To give an example, consider a random bipartite graph with
Poisson-distributed numbers of both movies per actor and actors per movie.
In this case, following the derivation of Eq.~\eref{poissongen}, we find
that
\begin{equation}
f_0(x) = \e^{\mu(x-1)},\qquad g_0(x) = \e^{\nu(x-1)},
\end{equation}
and $f_1(x)=f_0(x)$ and $g_1(x)=g_0(x)$.  Thus
\begin{equation}
G_0(x) = G_1(x) = \e^{\mu(\e^{\nu(x-1)}-1)}.
\end{equation}
This implies that $z_1=\mu\nu$ and $z_2=(\mu\nu)^2$, so that
\begin{equation}
\ell = {\log N\over\log \mu\nu} = {\log N\over\log z},
\end{equation}
just as in an ordinary Poisson-distributed random graph.  From
Eq.~\eref{biavs}, the average size $\av{s}$ of a connected component of
actors, below the phase transition, is
\begin{equation}
\av{s} = {1\over1-\mu\nu},
\end{equation}
which diverges, yielding a giant component, at $\mu\nu=z=1$, also as in the
ordinary random graph.  From Eqs.~\eref{giantcomp1} and~\eref{giantcomp2},
the size $S$ of the giant component as a fraction of $N$ is a solution of
\begin{equation}
S = 1 - \e^{\mu(\e^{-\nu S}-1)}.
\end{equation}
And from Eq.~\eref{bicc}, the clustering coefficient for the one-mode
network of actors is
\begin{equation}
C = {M \nu^3\over N \nu^2(\mu^2+\mu)} = {1\over \mu+1},
\end{equation}
where we have made use of Eq.~\eref{mMnN}.

Another quantity of interest is the distribution of numbers of co-stars,
i.e.,~of the numbers of people with whom each actor has appeared in a
movie.  As discussed above, this distribution is generated by the function
$G_0(x)$ defined in Eq.~\eref{defscg0}.  For the case of the Poisson degree
distribution, we can perform the derivatives, Eq.~\eref{derivatives}, and
setting $x=0$ we find that the probability $r_z$ of having appeared with a
total of exactly $z$ co-stars is
\begin{equation}
r_z = {\nu^z\over z!} \e^{\mu(\e^{-\nu}-1)}
      \sum_{k=1}^z \stirling{z}{k} \bigl[ \mu\e^{-\nu} \bigr]^k,
\label{defspz}
\end{equation}
where the coefficients $\bigl\lbrace{z\atop k}\bigr\rbrace$ are the
Stirling numbers of the second kind~\cite{AS65}
\begin{equation}
\stirling{z}{k} = \sum_{r=1}^k {(-1)^{k-r}\over r! (k-r)!}\,r^z.
\label{stirling}
\end{equation}

\subsection{Simulation results}
Random bipartite graphs can be generated using an algorithm similar to the
one described in Section~\ref{dirsim} for directed graphs.  After making
sure that the required degree distributions for both actor and movie
vertices have means consistent with the required total numbers of actors
and movies according to Eq.~\eref{mMnN}, we generate vertex degrees for
each actor and movie at random and calculate their sum.  If these sums are
unequal, we discard the degree of one actor and one movie, chosen at
random, and replace them with new degrees drawn from the relevant
distributions.  We repeat this process until the total actor and movie
degrees are equal.  Then we join vertices up in pairs.

\begin{figure}
\begin{center}
\psfig{figure=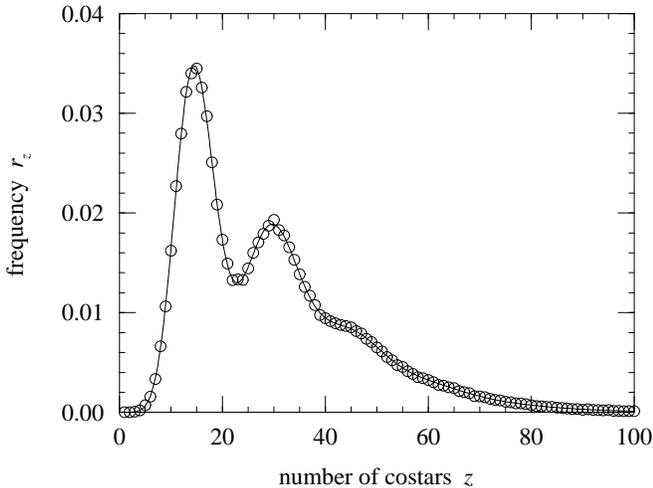,width=\columnwidth}
\end{center}
\caption{The frequency distribution of numbers of co-stars of an actor in
  a bipartite graph with $\mu=1.5$ and $\nu=15$.  The points are simulation
  results for $M=10\,000$ and $N=100\,000$.  The line is the exact
  solution, Eqs.~\eref{defspz} and~\eref{stirling}.  The error bars on the
  numerical results are smaller than the points.}
\label{pzdist}
\end{figure}

In Fig.~\ref{pzdist} we show the results of such a simulation for a
bipartite random graph with Poisson degree distribution.  (In fact, for the
particular case of the Poisson distribution, the graph can be generated
simply by joining up actors and movies at random, without regard for
individual vertex degrees.)  The figure shows the distribution of the
number of co-stars of each actor, along with the analytic solution,
Eqs.~\eref{defspz} and~\eref{stirling}.  Once more, numerical and analytic
results are in good agreement.

\section{Applications to real-world networks}
\label{applications}
In this section we construct random graph models of two types of real-world
networks, namely collaboration graphs and the world-wide web, using the
results of Sections~\ref{directed} and~\ref{bipartite} to incorporate
realistic degree distributions into the models.  As we will show, the
results are in reasonably good agreement with empirical data, although
there are some interesting discrepancies also, perhaps indicating the
presence of social phenomena that are not incorporated in the random graph.

\subsection{Collaboration networks}
\label{collabs}
In this section we construct random bipartite graph models of the known
collaboration networks of company directors~\cite{Davis96,DG97,DYB01},
movie actors~\cite{WS98}, and scientists~\cite{Newman01a}.  As we will see,
the random graph works well as a model of these networks, giving good
order-of-magnitude estimates of all quantities investigated, and in some
cases giving results of startling accuracy.

\begin{figure}
\begin{center}
\psfig{figure=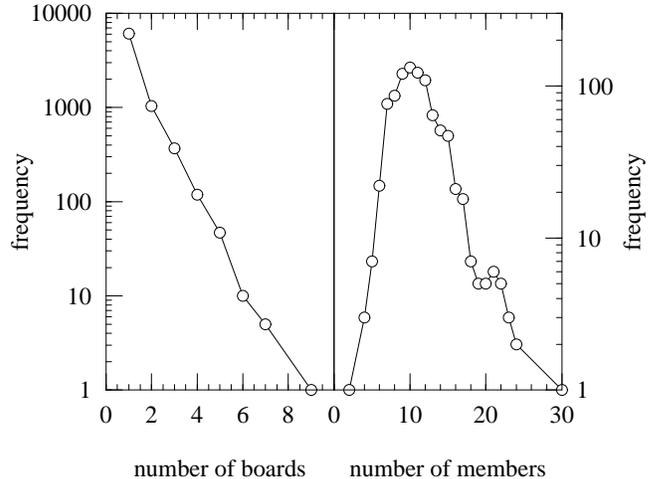,width=\columnwidth}
\end{center}
\caption{Frequency distributions for the boards of directors of the Fortune
  1000.  Left panel: the numbers of boards on which each director sits.
  Right panel: the numbers of directors on each board.}
\label{distmn}
\end{figure}

Our first example is the collaboration network of the members of the boards
of directors of the Fortune 1000 companies (the one thousand US companies
with the highest revenues).  The data come from the 1999 Fortune
1000~\cite{Davis96,DG97,DYB01} and in fact include only 914 of the 1000,
since data on the boards of the remaining 86 were not available.  The data
form a bipartite graph in which one type of vertex represents the boards of
directors, and the other type the members of those boards, with edges
connecting boards to their members.  In Fig.~\ref{distmn} we show the
frequency distribution of the numbers of boards on which each member sits,
and the numbers of members of each board.  As we see, the former
distribution is close to exponential, with the majority of directors
sitting on only one board, while the latter is strongly peaked around 10,
indicating that most boards have about 10 members.

\begin{table}[b]
\begin{center}
\begin{tabular}{l|cc|cc}
 & \multicolumn{2}{c|}{clustering $C$} & \multicolumn{2}{c}{average degree $z$} \\
 network                   & theory   & actual  & theory   & actual   \\
\hline
 company directors         & $0.590$  & $0.588$ & $14.53$  & $14.44$  \\
 movie actors              & $0.084$  & $0.199$ & $125.6$  & $113.4$  \\
 physics ({\tt arxiv.org}) & $0.192$  & $0.452$ & $16.74$  & $9.27$   \\
 biomedicine (MEDLINE)     & $0.042$  & $0.088$ & $18.02$  & $16.93$  \\
\end{tabular}
\end{center}
\caption{Summary of results of the analysis of four collaboration networks.}
\label{bitable}
\end{table}

Using these distributions, we can define generating functions $f_0(x)$ and
$g_0(x)$ as in Eq.~\eref{arbdist}, and hence find the generating functions
$G_0(x)$ and $G_1(x)$ for the distributions of numbers of co-workers of the
directors.  We have used these generating functions and Eqs.~\eref{biavz}
and~\eref{bicc} to calculate the expected clustering coefficient $C$ and
the average number of co-workers $z$ in the one-mode projection of board
directors on a random bipartite graph with the same vertex degree
distributions as the original dataset.  In Table~\ref{bitable} we show the
results of these calculations, along with the same quantities for the real
Fortune 1000.  As the table shows the two are in remarkable---almost
perfect---agreement.

\begin{figure}
\begin{center}
\psfig{figure=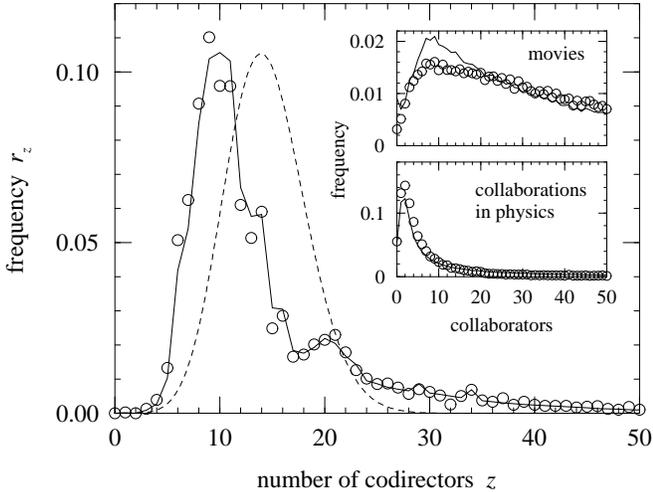,width=\columnwidth}
\end{center}
\caption{The probability distribution of numbers of co-directors in the
  Fortune 1000 graph.  The points are the real-world data, the solid line
  is the bipartite graph model, and the dashed line is the Poisson
  distribution with the same mean.  Insets: the equivalent distributions
  for the numbers of collaborators of movie actors and physicists.}
\label{directors}
\end{figure}

It is not just the average value of $z$ that we can calculate from our
generating functions, but the entire distribution: since the generating
functions are finite polynomials in this case, we can simply perform the
derivatives to get the probability distribution~$r_z$.  In
Fig.~\ref{directors}, we show the results of this calculation for the
Fortune 1000 graph.  The points in the figure show the actual distribution
of $z$ for the real-world data, while the solid line shows the theoretical
results.  Again the agreement is excellent.  The dashed line in the figure
shows the distribution for an ordinary Poisson random graph with the same
mean.  Clearly this is a significantly inferior fit.

\begin{figure}
\begin{center}
\psfig{figure=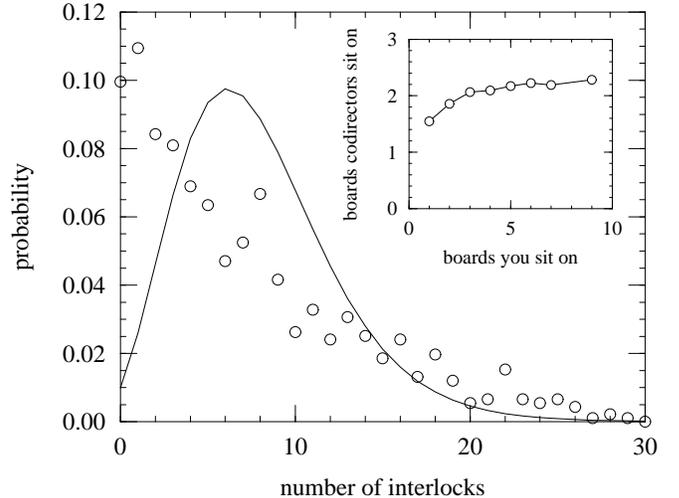,width=\columnwidth}
\end{center}
\caption{The distribution of the number of other boards with which each
  board of directors is ``interlocked'' in the Fortune 1000 data.  An
  interlock between two boards means that they share one or more
  common members.  The points are the empirical data, the solid line
  is the theoretical prediction.  Inset: the number of boards on which
  one's codirectors sit, as a function of the number of boards one
  sits on oneself.}
\label{interlocks}
\end{figure}

In fact, within the business world, attention has focussed not on the
collaboration patterns of company directors, but on the ``interlocks''
between boards, i.e.,~on the one-mode network in which vertices represent
boards of directors and two boards are connected if they have one or more
directors in common~\cite{Mariolis75,Davis96}.  This is also simple to
study with our model.  In Fig.~\ref{interlocks} we show the distribution of
the numbers of interlocks that each board has, along with the theoretical
prediction from our model.  As we see, the agreement between empirical data
and theory is significantly worse in this case than for the distribution of
co-directors.  In particular, it appears that our theory significantly
underestimates the number of boards which are interlocked with very small
or very large numbers of other boards, while over estimating those with
intermediate numbers of interlocks.  One possible explanation of this is
that ``big-shots work with other big-shots.''  That is, the people who sit
on many boards tend to sit on those boards with other people who sit on
many boards.  And conversely the people who sit on only one board (which is
the majority of all directors), tend to do so with others who sit on only
one board.  This would tend to stretch the distribution of numbers of
interlocks, just as seen in figure, producing a disproportionately high
number of boards with very many or very few interlocks to others.  To test
this hypothesis, we have calculated, as a function of the number of boards
on which a director sits, the average number of boards on which each of
their codirectors sit.  The results are shown in the inset of
Fig.~\ref{interlocks}.  If these two quantities were uncorrelated, the plot
would be flat.  Instead, however, it slopes clearly upwards, indicating
indeed that on the average the big-shots work with other big-shots.  (This
idea is not new.  It has been discussed previously by a number of
others---see Refs.~\onlinecite{MS85} and~\onlinecite{DM99}, for example.)

The example of the boards of directors is a particularly instructive one.
What it illustrates is that the cases in which our random graph models
agree well with real-world phenomena are not necessarily the most
interesting.  Certainly it is satisfying, as in Fig.~\ref{directors}, to
have the theory agree well with the data.  But probably
Fig.~\eref{interlocks} is more instructive: we have learned something about
the structure of the network of the boards of directors by observing the
way in which the pattern of board interlocks differs from the predictions
of the purely random network.  Thus it is perhaps best to regard our random
graph as a null model---a baseline from which our expectations about
network structure should be measured.  It is deviation from the random
graph behavior, not agreement with it, that allows us to draw conclusions
about real-world networks.

We now look at three other graphs for which our theory also works well,
although again there are some noticeable deviations from the random graph
predictions, indicating the presence of social or other phenomena at work
in the network.

We consider the graph of movie actors and the movies in which they
appear~\cite{WS98,note2} and graphs of scientists and the papers they write
in physics and biomedical research~\cite{Newman01a}.  In
Table~\ref{bitable} we show results for the clustering coefficients and
average coordination numbers of the one-mode projections of these graphs
onto the actors or scientists.  As the table shows, our theory gives
results for these figures which are of the right general order of
magnitude, but typically deviate from the empirically measured figures by a
factor of two or so.  In the insets of Fig.~\ref{directors} we show the
distributions of numbers of collaborators in the movie actor and physics
graphs, and again the match between theory and real data is good, but not
as good as with the Fortune 1000.

The figures for clustering and mean numbers of collaborators are
particularly revealing.  The former is uniformly about twice as high in
real life as our model predicts for the actor and scientist networks.  This
shows that there is a significant tendency to clustering in these networks,
in addition to the trivial clustering one expects on account of the
bipartite structure.  This may indicate, for example, that scientists tend
to introduce pairs of their collaborators to one another, thereby
encouraging clusters of collaboration.  The figures for average numbers of
collaborators show less deviation from theory than the clustering
coefficients, but nonetheless there is a clear tendency for the numbers of
collaborators to be smaller in the real-world data than in the models.
This probably indicates that scientists and actors collaborate repeatedly
with the same people, thereby reducing their total number of collaborators
below the number that would naively be expected if we consider only the
numbers of papers that they write or movies they appear in.  It would
certainly be possible to take effects such as these into account in a more
sophisticated model of collaboration practices.

\subsection{The world-wide web}
In this section we consider the application of our theory of random
directed graphs to the modeling of the world-wide web.  As we pointed out
in Section~\ref{dirgf}, it is not at present possible to make a very
accurate random-graph model of the web, because to do so we need to know
the joint distribution $p_{jk}$ of in- and out-degrees of vertices, which
has not to our knowledge been measured.  However, we can make a simple
model of the web by assuming in- and out-degree to be independently
distributed according to their known distributions.  Equivalently, we
assume that the joint probability distribution factors according to $p_{jk}
= p_j q_k$.

\begin{figure}
\begin{center}
\psfig{figure=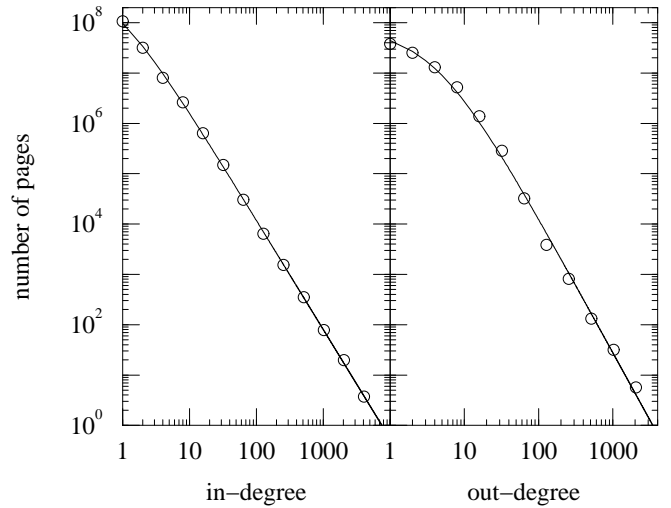,width=\columnwidth}
\end{center}
\caption{The probability distribution of in-degree (left panel) and out-degree
  (right panel) on the world-wide web, rebinned from the data of
  Broder~\etal~\protect\cite{BKMRRSTW00}.  The solid lines are best fits of
  form~\eref{curvy}.}
\label{broder}
\end{figure}

Broder~\etal~\cite{BKMRRSTW00} give results showing that the in- and
out-degree distributions of the web are approximately power-law in form
with exponents $\tau_{\rm in}=2.1$ and $\tau_{\rm out}=2.7$, although there
is some deviation from the perfect power law for small degree.  In
Fig.~\ref{broder}, we show histograms of their data with bins chosen to be
of uniform width on the logarithmic scales used.  (This avoids certain
systematic errors known to afflict linearly histogrammed data plotted on
log scales.)  We find both distributions to be well-fitted by the form
\begin{equation}
p_k = C (k+k_0)^{-\tau},
\label{curvy}
\end{equation}
where the constant $C$ is fixed by the requirement of normalization, taking
the value $1/\zeta(\tau,k_0)$, were $\zeta(x,y)$ is the generalized
$\zeta$-function~\cite{AS65}.  The constants $k_0$ and $\tau$ are found by
least-squares fits, giving values of $0.58$ and $3.94$ for $k_0$, and
$2.17$ and $2.69$ for $\tau$, for the in- and out-degree distributions
respectively, in reasonable agreement with the fits performed by
Broder~\etal\ \ With these choices, the data and Eq.~\eref{curvy} match
closely (see Fig.~\ref{broder} again).

Neither the raw data nor our fits to them satisfy the
constraint~\eref{inoutfix}, that the total number of links leaving pages
should equal the total number arriving at them.  This is because the data
set is not a complete picture of the web.  Only about 200 million of the
web's one billion or so pages were included in the study.  Within this
subset, our estimate of the distribution of out-degree is presumably quite
accurate, but many of the outgoing links will not connect to other pages
within the subset studied.  At the same time, no incoming links which
originate outside the subset of pages studied are included, because the
data are derived from ``crawls'' in which web pages are found by following
links from one to another.  In such a crawl one only finds links by finding
the pages that they originate from.  Thus our data for the incoming links
is quite incomplete, and we would expect the total number of incoming links
in the dataset to fall short of the number of outgoing ones.  This indeed
is what we see.  The totals for incoming and outgoing links are
approximately $2.3\times10^8$ and $1.1\times10^9$.

The incompleteness of the data for incoming links limits the information we
can at present extract from a random graph model of the web.  There are
however some calculations which only depend on the out-degree distribution.

Given Eq.~\eref{curvy}, the generating functions for the out-degree
distribution take the form
\begin{equation}
G_0(x) = G_1(x) = {\Phi(x,\tau,k_0)\over\zeta(\tau,k_0)},
\end{equation}
where $\Phi(x,y,z)$ is the Lerch $\Phi$-function~\cite{AS65}.  The
corresponding generating functions $F_0$ and $F_1$ we cannot calculate
accurately because of the incompleteness of the data.  The equality
$G_0=G_1$ (and also $F_0=F_1$) is a general property of all directed graphs
for which $p_{jk} = p_j q_k$ as above.  It arises because in such graphs
in- and out-degree are uncorrelated, and therefore the distribution of the
out-degree of a vertex does not depend on whether you arrived at it by
choosing a vertex at random, or by following a randomly chosen edge.

One property of the web which we can estimate from the generating functions
for out-degree alone is the fraction $S_{\rm in}$ of the graph taken up by
the giant strongly connected component plus those sites from which the
giant strongly connected component can be reached.  This is given by
\begin{equation}
S_{\rm in} = 1 - G_0(1-S_{\rm in}).
\end{equation}
In other words, $1-S_{\rm in}$ is a fixed point of $G_0(x)$.  Using the
measured values of $k_0$ and $\tau$, we find by numerical iteration that
that $S_{\rm in} = 0.527$, or about 53\%.  The direct measurements of the
web made by Broder~\etal\ show that in fact about 49\% of the web falls
in~$S_{\rm in}$, in reasonable agreement with our calculation.  Possibly
this implies that the structure of the web is close to that of a directed
random graph with a power-law degree distribution, though it is possible
also that it is merely coincidence.  Other comparisons between random graph
models and the web will have to wait until we have more accurate data on
the joint distribution $p_{jk}$ of in- and out-degree.

\section{Conclusions}
In this paper we have studied in detail the theory of random graphs with
arbitrary distributions of vertex degree, including directed and bipartite
graphs.  We have shown how, using the mathematics of generating functions,
one can calculate exactly many of the statistical properties of such graphs
in the limit of large numbers of vertices.  Among other things, we have
given explicit formulas for the position of the phase transition at which a
giant component forms, the size of the giant component, the average and
distribution of the sizes of the other components, the average numbers of
vertices a certain distance from a given vertex, the clustering
coefficient, and the typical vertex--vertex distance on a graph.  We have
given examples of the application of our theory to the modeling of
collaboration graphs, which are inherently bipartite, and the World-Wide
web, which is directed.  We have shown that the random graph theory gives
good order-of-magnitude estimates of the properties of known collaboration
graphs of business-people, scientists and movie actors, although there are
measurable differences between theory and data which point to the presence
of interesting sociological effects in these networks.  For the web we are
limited in what calculations we can perform because of the lack of
appropriate data to determine the generating functions.  However, the
calculations we can perform agree well with empirical results, offering
some hope that the theory will prove useful once more complete data become
available.

\section*{Acknowledgements}
The authors would like to thank Lada Adamic, Andrei Broder, Jon Kleinberg,
and Cris Moore for useful comments and suggestions, and Jerry Davis, Paul
Ginsparg, Oleg Khovayko, David Lipman, Grigoriy Starchenko, and Janet
Wiener for supplying data used in this study.  This work was funded in part
by the National Science Foundation, the Army Research Office, the Electric
Power Research Institute, and Intel Corporation.


\begin{references}
\bibitem{note1}
  Note that one must use $\log(1-r)$ in this expression, and not $\log r$,
  even though one might expect the two to give the same result.  The reason
  is that $r$ can be zero where $1-r$ cannot.  With the standard 32-bit
  random number generators used on most computers, $r$ will be zero about
  once in every 4 billion calls, and when it is, calculating $\log r$ will
  give an error but $\log(1-r)$ will not.  For simulations on large graphs
  of a few million vertices or more this will happen with some frequency,
  and calculation of $\log r$ should therefore be avoided.
\end{references}
\end{document}